\documentclass[]{elsart}

\usepackage{amssymb}
\usepackage{pst-node}
\usepackage[T1]{fontenc}
\usepackage[dvips]{graphicx}
\usepackage{latexsym}
\usepackage{epsfig}
\usepackage{rotate}
\usepackage{pstricks}
\usepackage[latin2]{inputenc}
\begin{document}
\begin{frontmatter}

\title{Simulation of Ultra-High Energy Photon Propagation in the Geomagnetic Field}

\author[a1]{P. Homola\corauthref{cor1}},
\ead{Piotr.Homola@ifj.edu.pl}
\author[a1]{D.Góra},
\author[a2]{D. Heck}, 
\author[a2]{H. Klages},
\author[a1]{J. Pękala}, 
\author[a2]{M. Risse},
\author[a1]{B. Wilczyńska},
\author[a1]{and H. Wilczyński}

\corauth[cor1]{ {\it Corresponding author}: tel.: +48 12 6628348, fax: +48 12 6628012
}
 
\address[a1]{Institute of Nuclear Physics PAS, Kraków,
ul. Radzikowskiego 152, 31-342 Kraków, Poland}

\address[a2]{Forschungszentrum Karlsruhe, Institut f\"ur Kernphysik, 76021 Karlsruhe,
Germany}







\begin{abstract}

The identification of primary photons 
or specifying stringent limits on the photon flux is of major importance for
understanding the origin of ultra-high energy (UHE) cosmic rays.
We present a new Monte Carlo program allowing detailed studies of conversion and 
cascading of UHE photons in the geomagnetic field. 
The program named PRESHOWER can be used both as an independent tool or 
together with a shower simulation code.
With the stand-alone version of the code it is possible to investigate
various properties of the particle cascade induced by UHE 
photons interacting in the Earth's magnetic field before entering
the Earth's atmosphere.   
Combining this program with an extensive air shower simulation code
such as CORSIKA offers the possibility of 
investigating signatures of photon-initiated showers. In particular,
features can be studied that help to discern such showers from the ones induced by hadrons.  
As an illustration, calculations for the conditions of the southern part of the Pierre Auger 
Observatory are presented. 
 
\end{abstract}


\end{frontmatter}

\section{Program Summary}
CPC Program Library Index: 1.1 Cosmic Rays\\
\begin{tabular}{p{7cm}p{7cm}}
Title of program: & PRESHOWER 1.0\\
Catalog number: &  \\
Program obtainable: & from P.~Homola, e-mail: Piotr.Homola@ifj.edu.pl \\
Computer on which the program has been thoroughly tested: & Intel-Pentium based PC \\
Operating system: & Linux, DEC-Unix \\
Programming languages used: & C, FORTRAN 77 \\
Memory required to execute: & $<100$ kB \\
No. of bits in a word: & 32 \\
Has the code been vectorized? & no \\
Number of lines in distributed program: & 1900 \\
Other procedures used in PRESHOWER: & IGRF~\cite{tsygan}, bessik, ran2~\cite{numrec} \\
Nature of physical problem: & simulation of a cascade of particles initiated by UHE photon 
passing through the geomagnetic field above the Earth's atmosphere.\\
\end{tabular}
\begin{tabular}{p{7cm}p{7cm}}
Method of solution: & The primary photon is tracked until its conversion into $e^+e^-$ pair
or until it reaches the upper atmosphere. If conversion occured each individual particle in
the resultant preshower is checked for either bremsstrahlung radiation (electrons) or 
secondary gamma conversion (photons). The procedure ends at the top of atmosphere and the
shower particle data are saved.\\
Restrictions on the complexity of the problem: & gamma conversion into particles other than electron 
pair hasn't been taken into account \\
Typical running time: & 100 preshower events with primary energy $10^{20}$ eV require a 800 MHz
CPU time of about 50 min., with $10^{21}$ eV the simulation time for 100 events grows up to 500 min.

\end{tabular}

\newpage
\section{Introduction}
The existence of ultra-high energy cosmic rays (UHECR) beyond the Greisen-Zatsepin-Kuzmin
(GZK) cutoff energy \cite{gzk} 
is one of the main puzzles in today's astrophysics.
The composition of UHECR, if known, could provide a stringent test of models
of particle acceleration and propagation towards the Earth.
Most theoretical efforts are made towards explaining the acceleration of
protons to ultra-high energies. This is because protons seem to be
the most promising candidates for the UHECR particles. There are, however, also so-called exotic
models which postulate decays of supermassive ``X-particles''. A common prediction
of these models is that there should be a significant photon flux 
in the ultra-high energy region. 
Accurate identification of photon primaries, measurement of the ultra-high energy
photon flux, or
specifying the upper limit for it, will provide an excellent test for the models 
of cosmic-ray origin.

Ultra-high energy photons in the presence of the geomagnetic field can convert into 
an electron-positron
pair before they enter the atmosphere.
Synchrotron radiation of this pair gives rise to a cascade of photons.
Some of them can have energies sufficient for the next conversion. 
The cascade at the top of the atmosphere thus consists mainly of
photons and two or more electrons.
We will call this cascade a ``preshower'' since it originates and develops 
above the upper atmosphere, i.e. before the ``ordinary'' shower development in the air. 

Particles of a preshower induce subshowers after entering the atmosphere.
A superposition of these subshowers will be seen by detectors as one extensive
air shower, whose profile is expected to differ from the profile of a shower
induced by a single unconverted photon. 

Forthcoming experiments with large collecting apertures 
(Pierre Auger Observatory \cite{auger}, EUSO \cite{euso}),
that are designed to detect cosmic rays 
of ultra-high energies,
give a good opportunity to test these models.
A photon-induced shower might have observable features that allow to distinguish it 
from showers initiated by hadrons.
 
In this paper we describe the PRESHOWER program that allows 
extensive studies of properties of 
photon-induced showers. The program performs a simulation of the propagation of
photons before they enter the Earth's atmosphere and can be used both as an independent 
analysis tool as well as a part of an air shower simulation package 
(e.g. CORSIKA \cite{corsika}). 

With the PRESHOWER program working in stand-alone mode we performed, as an illustration,
an analysis of basic
properties of preshowers for the magnetic conditions of Malarg\"ue in Argentina -- the site
of the Southern Pierre Auger Observatory~\cite{auger}.
The results of these studies, which are
presented in Sec. \ref{results}, are in general agreement with other analyses 
\cite{mcbr,karakula,stanev,bedn,vankov1,vankov2,billoir}. 

By linking our program to an air shower
simulation code one is able to study showers initiated by various primaries.
We checked this by combining our PRESHOWER code with the CORSIKA air shower simulation 
package which simulates air showers initiated by the preshower particles.
We simulated photon-induced 
showers of different energies
and different directions at the Southern Pierre Auger Observatory. 
Although the interaction in the geomagnetic field changes significantly
the shower properties so that primary photon events look more similar
to hadronic showers, a discrimination between primary photons and hadrons
still seems possible.

The plan of the paper is the following: The relevant effects and the formulas
applied in the simulation are summarized in Section~\ref{prephys}.
The structure of the program is discussed in Section~\ref{structure}.
As an illustration, in Section~\ref{results} results obtained with the code for
the case of the Auger Observatory are discussed. 
The paper is concluded in Section~\ref{summary}.

\section{Preshower physics}
\label{prephys}
At energies above 10 EeV, in the presence of the geomagnetic field, a photon can convert
into an $e^+e^-$ pair before entering the atmosphere. 
The resultant electrons will subsequently
lose their energy by magnetic bremsstrahlung. The emitted photons can convert again
if their energy is high enough. In this way, instead of one high
energy photon at the top of atmosphere, a number of less energetic
particles, mainly photons and a few electrons, will emerge constituting a preshower.
A superposition of subshowers induced by these particles
should be seen by detectors as one extensive air shower which
reaches its maximum earlier than a shower induced by a single photon of the same primary
energy. 
This is also due to the Landau-Pomeranchuk-Migdal (LPM) 
effect \cite{lpm} which is stronger for photons of higher energies. 
Both photon conversion and magnetic bremsstrahlung are strongly dependent
on the transverse component of the geomagnetic field.
In the following, we concentrate on these effects, give the main formulas
and discuss their application.

\subsection{Geomagnetic field model}
\label{igrf}
Since the geomagnetic field is responsible both for gamma conversion and for
bremsstrahlung of electrons, it is very important to model it with
adequate accuracy. We use the International Geomagnetic
Reference Field model (IGRF)~\cite{igrf} for $n=10$, $n$ being the
highest order of spherical harmonics in the scalar potential.
To calculate the
field according to the IGRF model we use the numerical procedures written by
Tsyganenko~\cite{tsygan}.
When choosing $n=1$, the output of these procedures
is equivalent to a dipole model with $|{\bf M}|=8\cdot10^{25}$ G$\cdot$cm$^3$ and with poles
at 79.3$^\circ$N, 71.5$^\circ$W and 79.3$^\circ$S, 108.5$^\circ$E~\cite{igrf}.

\subsection{Magnetic pair production: $\gamma \rightarrow e^+e^-$}
\label{magneticpp}
The number of pairs created by a high-energy photon in the presence of a magnetic
field per path length $dr$
can be expressed in terms of the attenuation coefficient $\alpha(\chi)$ \cite{Erber}:
\begin{equation}
\label{npairs}
n_{pairs}=n_{photons}\{1-\exp[-\alpha(\chi)dr]\}
\end{equation}
where
\begin{equation}
\label{alpha}
\alpha(\chi)=0.5(\alpha_{em} m_ec/\hbar)(B_\bot/B_{cr})T(\chi)
\end{equation}
with $\alpha_{em}$ being the fine structure constant, $\chi\equiv0.5(h\nu/m_ec^2)(B_\bot/B_{cr})$,
$B_\bot$ is the magnetic field component transverse to the direction of the photon's motion,
$B_{cr}\equiv m_e^2c^3/e\hbar=4.414\times 10^{13}$ G
and $T(\chi)$ is the magnetic pair production function. $T(\chi)$ can be well approximated by:
\begin{equation}
\label{tcentral}
T(\chi)\cong0.16\chi^{-1}{K^2}_{1/3}(\frac{2}{3\chi}),
\end{equation}
where $K_{1/3}$ is the modified Bessel function of order $1/3$. For small or large arguments
$T(\chi)$ can be approximated by

\begin{equation}
\label{tlimits}
\begin{array}{c}
T(\chi)\cong\left\{ \begin{array}{ll}
0.46\exp(-\frac{4}{3\chi}), & ~~\chi \ll 1;\\
0.60\chi^{-1/3}, & ~~\chi \gg 1.
\end{array} \right.
\end{array}
\end{equation}
Representative values of $T(\chi)$ are listed in Table \ref{Ttab}.

We use Eq.~(\ref{npairs})
to calculate the probability of $\gamma$ conversion over a small path length $dr$:
\begin{equation}
\label{pconv}
p_{conv}(r)=1-\exp[-\alpha(\chi(r))dr]\simeq\alpha(\chi(r))dr
\end{equation}
and the probability of $\gamma$ conversion over a longer distance $R$:
\begin{equation}
\label{pconv_int}
P_{conv}(R)=1-\exp[-\int^R_0\alpha(\chi(r))dr]
\end{equation} 

\subsection{Magnetic bremsstrahlung}
After photon conversion, the electron-positron pair propagates.
The energy distribution in an $e^+e^-$ pair
is computed according to Ref. \cite{ppdaugherty}:

\begin{equation}
\frac{d\alpha(\varepsilon,\chi)}{d\varepsilon}\approx\frac{\alpha_{em}m_ec B_\bot}{\hbar B_{cr}}
\frac{3^{1/2}}{9\pi\chi}\frac{[2+\varepsilon(1-\varepsilon)]}{\varepsilon(1-\varepsilon)}
K_{2/3}\left[\frac{1}{3\chi\varepsilon(1-\varepsilon)}\right],
\label{daug}
\end{equation}
where $\varepsilon$ denotes the fractional energy of an electron and the other symbols
were explained in the previous chapter. 
The probability of asymmetric energy partition grows with the primary photon energy and 
with the magnetic field. Beginning from $\chi>10$, the asymmetric energy partition is 
even more favored than the symmetric one.

\begin{figure}
\begin{center}
\includegraphics[width=0.7\textwidth,angle=0]{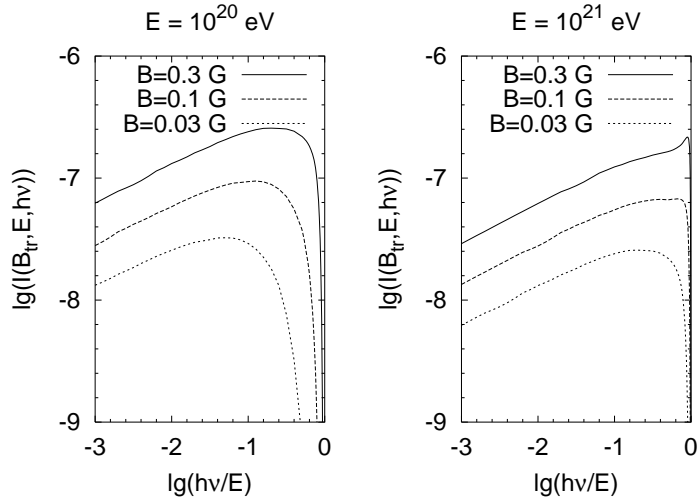}
\end{center}
\caption {Bremsstrahlung spectral distribution: $E=10^{20}$ eV (left)
and $E=10^{21}$ eV (right) for different transverse magnetic field strengths: $B=0.3$ G,
$0.1$ G and $0.03$ G for the uppermost, middle and lowest curves 
respectively.}
\label{sokolov}
\end{figure}

Electrons traveling at relativistic speeds in the presence of a magnetic field
emit bremsstrahlung radiation (synchrotron radiation). For electron energies
$E \gg m_ec^2$ and for $B_\bot \ll B_{cr}$, the spectral distribution of radiated
energy is given in Ref. \cite{sokolov}:
\begin{equation}
f(y)=\frac{9\sqrt{3}}{8\pi}\frac{y}{(1+\xi y)^3}\left\{\int^\infty_yK_{5/3}(z)dz+
\frac{(\xi y)^2}{1+\xi y}K_{2/3}(y)\right\}
\label{fy}
\end{equation}
where $\xi =(3/2)(B_\bot/B_{cr})(E/m_ec^2)$, $E$ and $m_e$ are electron initial energy
and rest mass respectively, $K_{5/3}$ and $K_{2/3}$ are modified Bessel functions,
and $y$ is related to the emitted photon energy $h\nu$ by
\begin{equation}
y(h\nu)=\frac{h\nu}{\xi (E-h\nu)} \;; \qquad \qquad
\frac{dy}{d(h\nu)}=\frac{E}{\xi(E-h\nu)^2}
\label{yhv}
\end{equation}
The total energy emitted per unit distance is (in CGS units)
\begin{equation}
W=\frac{2}{3}r_0^2B_\bot^2\left(\frac{E}{m_ec^2}\right)^2\int^\infty_0f(y)dy
\label{W}
\end{equation}
with $r_0$ being the classical electron radius. For our purposes we use the spectral distribution
of radiated energy defined as
\begin{equation}
I(B_\bot,E,h\nu)\equiv\frac{h\nu dN}{d(h\nu)dx}~~,
\label{Idef}
\end{equation}
where $dN$ is the number of photons with energy between $h\nu$ and $h\nu+d(h\nu)$
emitted over a distance $dx$. From Eqs.~(\ref{fy}), (\ref{yhv}), (\ref{W}), 
and (\ref{Idef})
we get \footnote{
Expression (\ref{brem}), valid for all values of $h\nu$, is equivalent to Eq. (2.5a)
in Ref. \cite{Erber}. A simplified form of distribution (\ref{brem}) is given by Eq. (2.10)
in Ref. \cite{Erber}, however it can be used only for $h\nu \ll E$.}
\begin{equation}
I(B_\bot,E,h\nu)=\frac{2}{3}r_0^2B_\bot^2\left(\frac{E}{m_ec^2}\right)^2f(h\nu)\frac{E}
{\xi(E-h\nu)^2}~~.
\label{brem}
\end{equation}
Provided $dx$ is small enough, $dN$ can be interpreted as a probability
of emitting a photon of energy between $h\nu$ and $h\nu+d(h\nu)$ by an electron
of energy $E$ over a distance $dx$. In our simulations we use a small step size of $dx=1$ km.
The total probability of emitting a photon in step $dx$ can then be written as
\begin{equation}
P_{brem}(B_\bot,E,h\nu,dx)=\int dN=dx\int^E_0 I(B_\bot,E,h\nu)\frac{d(h\nu)}{h\nu}~~.
\label{bremprob}
\end{equation}
The shape of the bremsstrahlung spectrum for two different energies is
shown in Fig. \ref{sokolov}. The three curves in each plot correspond to different values
of the transverse magnetic field: 0.3 G, 0.1 G and 0.03 G from the uppermost to the lowest
curve respectively.
The energy of the emitted photon is simulated according to the
probability density distribution $dN/d(h\nu)$ obtained from Eq.~\ref{bremprob}.

\subsection{Related physics topics}
In this section we shortly discuss the importance of other physics effects
connected to our main subject. These are:
\begin{itemize}
\item The geomagnetic field deflects the electrons from the initial preshower direction.
To evaluate this deflection we use the equation
\begin{equation}
\label{lorentz}
\frac{m_{rel}v^2}{R}=evB_\bot \;; \qquad m_{rel}=\frac{E}{c^2}
\end{equation}
where $v$ is the speed of electron and $e$ is the
elementary charge, to find the radius of curvature $R$.
If $R$ is much larger than the electron path length $L$ and 
assuming constant values of $B_\bot$, $E$ and $v \approx c$,
the linear displacement from the preshower core in the plane
perpendicular to the initial direction is well approximated by:
\begin{equation}
\label{dx}
\Delta x\cong\frac{L^2}{2R}=\frac{ecB_\bot L^2}{E}
\end{equation}
For typical values like $B_\bot=0.1$ G, $L=1000$ km
(see Section~\ref{results}) and relatively small energy of $E=10^{18}$ eV,
we get $R \approx 10^{11}$ km which gives $\Delta x \cong 0.5$ cm.
Assuming a larger value for $L$,
one should keep in mind that at
higher altitudes ${\bf B}$ is lower and $E$ is higher.
Allowing for other possible combinations of $B_\bot$, $E$ and $L$
one comes to a very safe conclusion 
that in any case $\Delta x <$ 1 m.
\item The initial angular spread of particles in both pair production and brems\-strah\-lung 
can be approximated by $\Delta\theta\simeq m_e/E$ \cite{deflect}. Even for energies
as low as $10^{18}$ eV one has $\Delta\theta\approx10^{-12}$ which even
for extremely long paths like for instance 10000 km results in negligible linear deviations
at the top
of the atmosphere of $\Delta x\approx 10^{-3}cm$.
\item A $\gamma$-ray traveling through the vicinity of the Sun 
can cascade in  the magnetosphere of the Sun where the field is much stronger 
than near the Earth.
The particles originated in conversion near the Sun can be significantly deflected 
from their initial direction by the solar magnetic field. 
The cascade of such particles after reaching the Earth's atmosphere
will induce an air shower of large lateral extent which might be well distinguishable
from proton or iron showers. 
However, the predicted rate of such events for primary energies of order $10^{19}$ eV is
very low: about 1 event per 10 years for the Pierre Auger Observatory \cite{bednarek}.
In our simulations we do not consider cascading in the Sun's magnetosphere.
\item The magnetic field embedded in the solar wind is of order $10^{-5}$ G \cite{spaceweather}
and it influences the Earth's magnetosphere at altitudes of order 8-10$R_\oplus$. 
Even for a primary energy as high as $10^{21}$ eV
gamma conversion takes place within
2$R_\oplus$ above the sea level (see Section~\ref{results})
where the Earth magnetic field is not less than $10^{-2}$ G.
Thus, the solar wind can be neglected in our studies. 
\item The preshower electrons traveling at speeds $v<c$ are delayed with respect
to the photons. This delay however is very small and can be neglected. 
For an extreme case of a $10^{18}$ eV electron traveling a path of
20000 km, the delay is of order $10^{-26}$~s.
The delay of the particles not parallel to the main direction is also negligible (see
the above discussion on the lateral spread of a preshower).
\end{itemize}

\section{Structure of the program}
\label{structure}
The following subsections refer to the stand-alone mode of the PRESHOWER program.
The PRESHOWER-CORSIKA mode requires only the main procedure
generating events, \texttt{preshw}, which is described in subsection \ref{generation}. 
All the input data in the PRESHOWER-CORSIKA mode are provided via the main 
CORSIKA input file, the standard output is printed within the CORSIKA output, 
and the \texttt{multirun.dat} file is not returned, since there is only
one run per one primary trajectory allowed in CORSIKA. There is also a different random number
generator used in the PRESHOWER-CORSIKA mode than in the stand-alone one. 
For more information on the PRESHOWER-CORSIKA mode the user can refer to the latest CORSIKA manual 
\cite{corsika_manual} or contact the corresponding author.
 
\subsection{The source code and other files in the PRESHOWER packet}
The PRESHOWER code consists of the following files:
\begin{description}
\item[preshw.c] contains the main procedure generating preshowers, \texttt{preshw}, 
and the other 
procedures in C that are called by \texttt{preshw}. 
\item[prog.c] is used only in the stand-alone mode. It reads the input 
file \texttt{INPUT} and calls the preshower procedure \texttt{preshw}. 
\item[igrf.f] is an external procedure for calculation of the geomagnetic field according to the 
IGRF model (author: N.A.Tsyganenko \cite{tsygan}), it is called by \texttt{preshw}.
\item[rmmard.f] is a dummy Fortran routine used only in the stand-alone version for
the proper linking of the program. 
In the PRESHOWER-CORSIKA mode the actual \texttt{rmmard} is responsible
for random number generation. 
\end{description}
For the user's convenience there are some more files included in the PRESHOWER packet:
\begin{description}
\item[Makefile] compilation and linking commands under linux. 
\item[README.txt] short information on the content of the PRESHOWER packet and on the 
installation process.
\item[INPUT] typical input data read by the program working in the stand-alone mode. In the
PRESHOWER-CORSIKA mode all the input data required by \texttt{preshw} are read directly from
the main CORSIKA input file.
\item[part\_out.dat] the example of a detailed output file containing information on the 
preshower particle data being the result of running the program with the input \texttt{INPUT}.  
\item[multirun.dat] the example of an output file summarizing the program run with the 
input \texttt{INPUT}; this file is produced only in the stand-alone mode. 
\item[out.txt] the standard output produced by the program in the stand-alone mode with
the use of the input data from \texttt{INPUT}. 
\end{description}

\subsection{Input and output data}
\label{inout}
The input data are read by the main program from the file \texttt{INPUT} and then 
passed to the \texttt{preshw} procedure. These data are
\begin{enumerate}
\item the primary photon energy, [GeV]
\item the zenith angle of the shower trajectory in the local reference frame (see Sec. \ref{results}), [deg]
\item the azimuth angle of the shower trajectory in the local frame, [deg]
\item the top of the atmosphere -- e.g. the value obtained from CORSIKA, [km a.s.l.]
\item geographical position of the observatory: longitude [deg]
\item geographical position of the observatory: latitude [deg]
\item the year of observation (needed for calculation of the field)
\item number of runs
\end{enumerate}
An exemplary set of the input data can be found in the file \texttt{INPUT}.

There are two output files produced by the PRESHOWER program: \texttt{part\_out.dat} 
and \texttt{multirun.dat}. Both files begin with a header containing all the crucial run
information, then comes the description of columns and then the data.
The data in \texttt{part\_out.dat} are presented
in two columns: particle id (1 - photons, 2 - positrons, 3 - electrons)
and particle energy at the top of the atmosphere (in eV). The
columns in \texttt{multirun.dat} are explained as follows:
\begin{enumerate}
\item the run number 
\item altitude of the first photon conversion in km a.s.l. 
\item total number of particles at the top of the atmosphere 
\item number of photons 
\item number of electrons 
\item maximum electron energy 
\item maximum photon energy 
\item fraction of energy carried by electrons 
\item total energy carried by preshower particles = primary photon energy 
\item total energy carried by electrons 
\item total energy carried by photons 
\end{enumerate}
The information printed as standard output begins with the same header as the output files
and is followed by the specific information on each run (all energies given in eV): 
the run number, the altitude of each conversion together with the 
energy partition information, preshower summary at the top of the atmosphere: 
total energy carried by photons (Etot\_g), total energy carried by electrons (Etot\_e), 
maximum photon energy (Eg\_max), maximum electron energy (Ee\_max),
number of particles (n\_part), number of photons (n\_phot), number of positrons (n\_e+) and
number of electrons (n\_e-). The example of a standard output is the file \texttt{out.txt}
included in the PRESHOWER packet.

\subsection{Generation of events}
\label{generation}
After reading the input parameters (see Sec.\ref{inout}) the event generation procedure
\texttt{preshw} is called, within which all the calculations are performed.
In the initial stage all the necessary parameter transformations are applied and 
the preshower trajectory vector is computed. 
The propagation path length is set to be $5R_\oplus$ long, which e.g.
for vertical showers means the simulation starting altitude of $5R_\oplus$ above sea level. 
All the data concerning the particles in the preshower are stored in the array 
\texttt{part\_out} which
at the beginning has only one entry: the primary photon. In steps of
$10$~km the transverse magnetic field is computed by the procedure \texttt{B\_transverse}, 
then the probability of conversion is found using Eq.~(\ref{pconv}).
The step size has been checked to be sufficiently small.
If the conversion takes place, 
the particle array is updated by replacing the photon with an $e^+$ and
adding an $e^-$ at the array end. The energy partition fraction is found by the
function \texttt{ppfrac}.
To assure accurate bremsstrahlung
simulation, beginning from the first gamma conversion the step of 1 km is used
for all particle types. In every step, for each photon in the array,
we check again for the conversion effect. In case of electrons the bremsstrahlung
radiation is simulated using function \texttt{brem}. The probability of emitting
a photon is calculated from Eq.~(\ref{bremprob}) with the integration 
starting from $h\nu=10^{12}$ eV 
and logarithmic steps of 0.01.
Bremsstrahlung photons of energies lower than $10^{12}$ eV have
a very small influence on the air shower evolution 
(see also Section~\ref{results}) and hence they are neglected.

Once a bremsstrahlung photon is emitted, its energy is determined by the probability
distribution $dN/d(h\nu)$ obtained again from Eq. (\ref{bremprob}), and the energy of
the radiating electron is diminished correspondingly. Simulations are finished
when the (adjustable) altitude of the top of the atmosphere is reached
and all preshower particle parameters
are written out (in the stand-alone mode) or passed to CORSIKA.

\subsection{Compilation, linking and running the program}

In order to compile, link and run PRESHOWER under Linux one should first
create a file \texttt{nr\_fun.c} in the PRESHOWER home directory and place the 
header: 

\texttt{\#include <stdio.h>}\\
\texttt{\#include <math.h>}

at the beginning of this file.
Then, below the header, the source code of the Numerical Recipes \cite{numrec}
procedures
\texttt{ran2} and \texttt{bessik} (together with the procedures called by 
\texttt{bessik}: \texttt{beschb}, \texttt{chebev} and \texttt{nrerror}) should be pasted. 
To compile and link the program one just types ``make'' in the PRESHOWER
home directory. For Unix compilation and linking one needs to replace the proper commands
in Makefile. Once the data in the \texttt{INPUT} file are appropriate, the program 
can be started by typing ``./preshower''. The compilation and linking process can be verified
by running the program with the exemplary \texttt{INPUT} file included in the packet. If the 
program runs correctly, it should return the same output files as those included in the packet
(\texttt{part\_out.dat}, \texttt{multirun.dat}) and the standard output as in \texttt{out.txt}.
 
\section{Illustration of preshower simulations}
\label{results}

The results presented in the following are obtained for the
magnetic conditions of the Southern Auger 
Observatory in Malarg\"ue, Argentina (35.2$^\circ$S, 69.2$^\circ$W).
Other geographical positions could easily be adopted.
The shower trajectories are given in the local frame
where the azimuth increases in the counter-clockwise direction and
$\phi=0^\circ$ means a shower coming from the geographical North.

To visualize how the transverse magnetic field ($B_\bot$) 
can change along the shower trajectory, we
have chosen some
representative directions along which a particle encounters
$B_\bot$ of significantly different strengths: strong,
medium and weak. 
The ``strong field direction'' is the one along which
$B_\bot$ is maximal compared to other directions with zenith angles $\theta\leq 60^\circ$
and is defined by the angles $\theta=60^{\circ}$ and $\phi=177^{\circ}$.
Similarly, the weak $B_\bot$ trajectory (``weak field direction'')
was chosen to resemble
the local field direction at Malarg\"ue of $\theta=50^\circ$, $\phi=357^\circ$ \cite{igrf}.
The ``medium field direction''
is a vertical trajectory ($\theta=0^{\circ}$).
In the following we will refer to these directions.

\subsection{Geomagnetic cascading}
\begin{figure}
\begin{center}
\includegraphics[width=0.7\textwidth,angle=0]{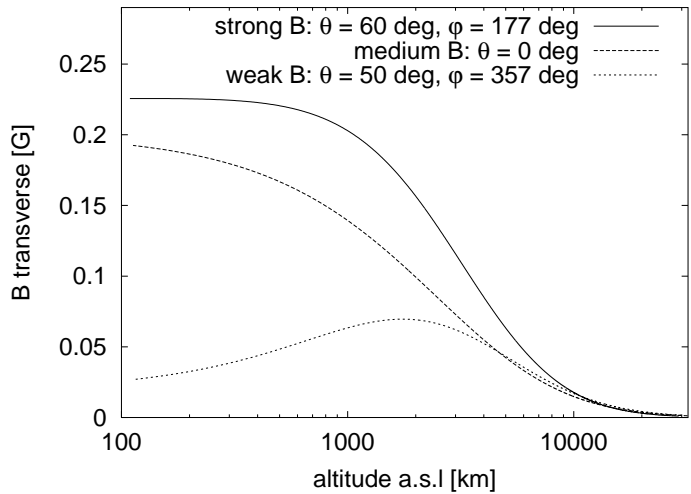}
\end{center}
\caption {$B_\bot$ along different shower trajectories for the
southern Auger Observatory.
Curves from top to bottom:
strong, medium, and weak field direction (see text).}
\label{fields}
\end{figure}

In Fig.~\ref{fields} we show how 
$B_\bot$ changes with altitude for the strong, medium and 
weak field directions.
The trajectories end at the chosen top of the atmosphere.
In addition to the altitude dependence of $B_\bot$,
large directional differences are visible.
Of course we expect the strong field direction
to coincide with the direction of a stronger preshower effect.

\begin{figure}
\begin{center}
\includegraphics[width=0.7\textwidth]{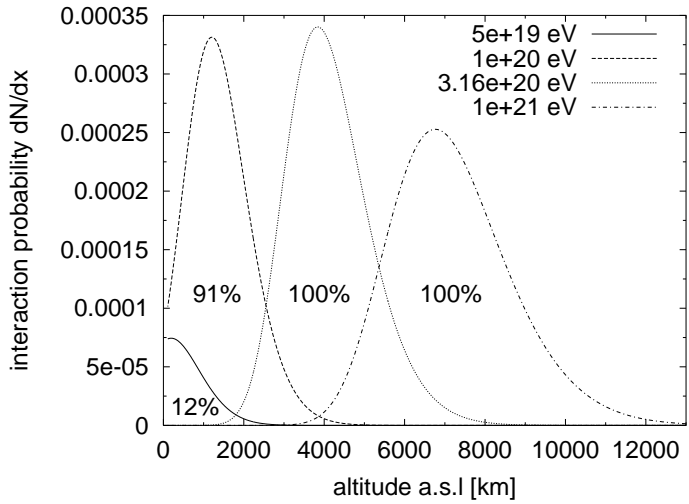}
\end{center}
\caption {Altitudes of first conversion along 
strong field direction
at Malarg\"ue plotted for four different primary photon 
energies. The numbers attached to the curves indicate the total conversion
probability (see text).}
\label{fpp}
\end{figure}

In Fig. \ref{fpp}, the altitudes of first photon conversion
are shown for different energies in the strong field direction.
The percentage values shown for each curve denote the total conversion probability
up to the top of atmosphere, which is adjusted here to the starting altitude
of CORSIKA at 112 km. It can be seen that  
for the magnetic field above Malarg\"ue,
the pair production effect becomes important for energies above
$5\times10^{19}$ eV. For all energies of interest, even in case of photons at $10^{21}$ eV, 
the conversion altitude is not higher than 13000 km, i.e. about $2R_\oplus$ a.s.l. 
Thus, the range of interest of the magnetic field strength for the case of Malarg\"ue
is $0.3$ G $\div$ $0.01$ G (see~Fig.~\ref{fields}).

\begin{figure}
\begin{center}
\includegraphics[width=0.7\textwidth]{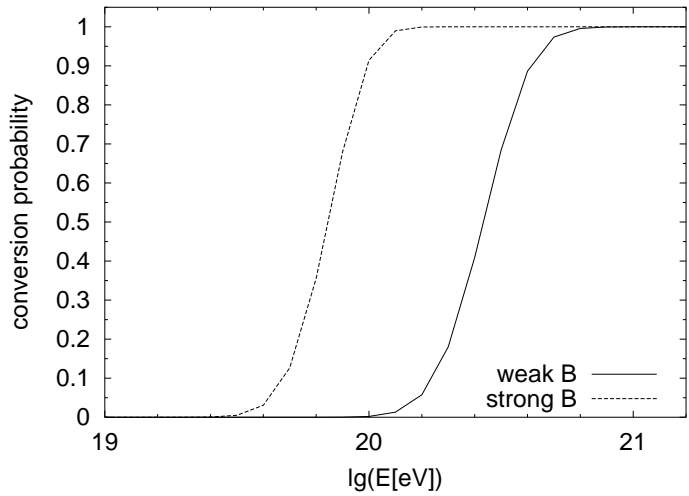}
\end{center}
\caption {Integrated conversion probability vs initial $\gamma$ energy
for weak and strong field directions.}
\label{conv_vs_E}
\end{figure}

The distribution plotted in Fig. \ref{fpp} is consistent with the results
of Ref. \cite{stanev}. The integrated conversion probability computed with 
Eq.~(\ref{pconv_int}) for different $\gamma$ energies and for the two trajectories of
strong and weak field directions
is plotted in Fig. \ref{conv_vs_E}. 
For photons of energies between $5\times10^{19}$ eV and $5\times10^{20}$ eV
a change in the arrival direction, i.e. a change in the magnetic
field component perpendicular to the photon trajectory,
may imply a dramatic increase or decrease
of the conversion probability. Thus, for these energies we expect 
a strong directional dependence of the preshower effect which
might serve as characteristic fingerprint of photons as primary cosmic rays.

\begin{figure}
\begin{center}
\includegraphics[width=1.0\textwidth]{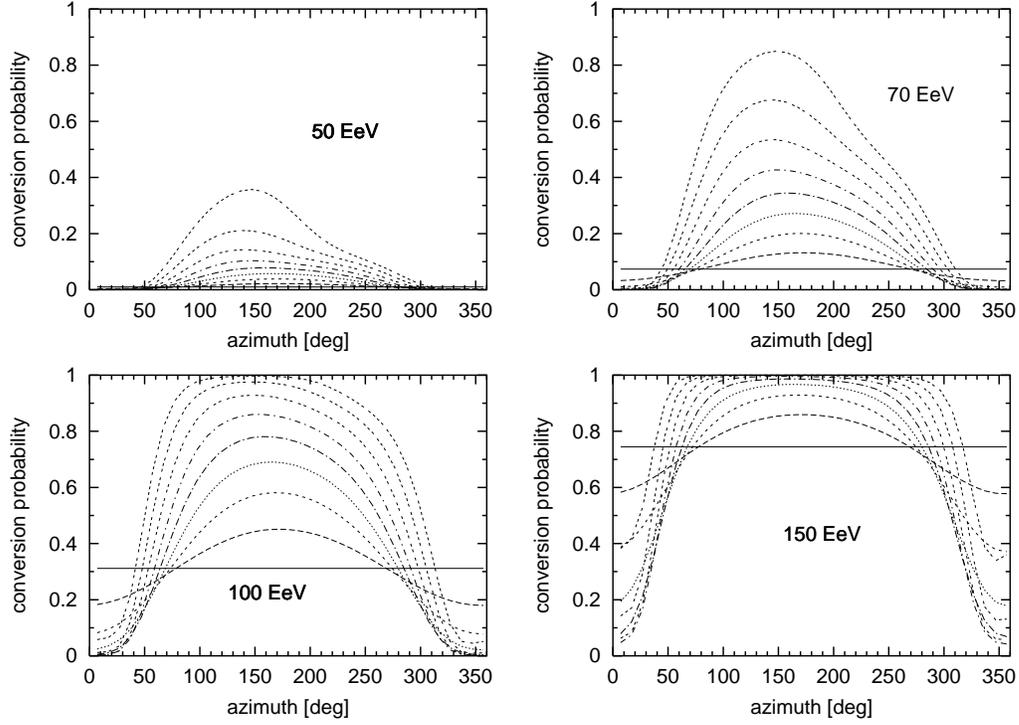}
\end{center}
\caption {
Total probability of $\gamma$ conversion for different arrival directions
and four initial energies. Each curve corresponds to a different zenith angle: 
$\theta=80^\circ$ for the uppermost curve down to
$\theta=0^\circ$ for the lowest one in steps of $10^\circ$.
Computations are made for Malarg\"ue magnetic conditions and
azimuth $0^\circ$ means the shower arriving from the geographic North.}
\label{maps}
\end{figure}

The directional dependence of the gamma conversion probability is shown in more
detail in Fig. \ref{maps} 
for different arrival directions and primary energies. 
The conversion effect starts to be important for directions towards the magnetic
South pole at photon energies 
around 50 EeV. For energies of about 150 EeV the conversion is very probable
for almost every direction.   
The photons coming from azimuthal angles around 150$^\circ$  
at large zenith angles convert more likely
than the others. To explain this, one notes that
for a given altitude the magnetic field is strongest around the geomagnetic axis and
of course the field becomes stronger with decreasing altitude.  
Since the photons coming from the geomagnetic South, 
which corresponds to the azimuth about 150$^\circ$ in case of Malarg\"ue,
cross the very vicinity of the geomagnetic axis, they must encounter
the stronger magnetic field than the others and hence their conversion
probability is higher. The larger zenith angle for these photons means that 
the strong field region is crossed at the lower altitudes where the field is stronger
and the conversion process is more effective. 
The plots in Fig. \ref{maps} are consistent with the values given in Ref. \cite{billoir}. 
 
All the preshowers discussed here are presented in their
final stage of development at the ``top of the atmosphere'' which in our simulations is 
assumed to be 112 km a.s.l. -- where in the CORSIKA code shower simulations are started.

\begin{figure}
\begin{center}
\includegraphics[width=1.0\textwidth,angle=0]{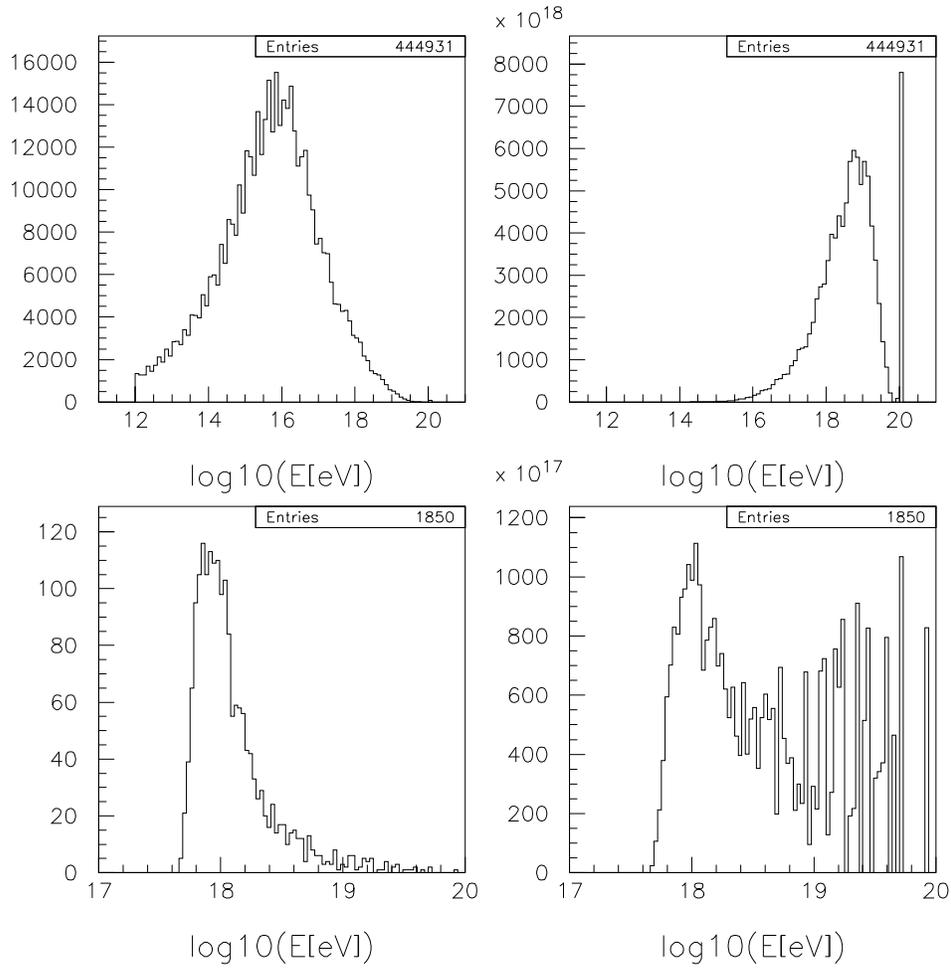}
\end{center}
\caption {Energy distribution of photons (top) and electrons (bottom) in 1000 preshowers initiated by
$10^{20}$ eV photons arriving at Malarg\"ue from the strong field
direction, compare Table \ref{pre_examples}. Spectra weighted by energy are plotted to the right.}
\label{20strong_profiles}
\end{figure}

In Fig.~\ref{20strong_profiles} mean energy spectra are plotted for 1000 preshowers 
initiated by photons of energy $10^{20}$ eV arriving at the top of atmosphere
from the strong field direction for Malarg\"ue. 
In the top row we show the distribution of photons (left) and the distribution
of photons weighted by their energies (right). 
We start the simulations of the bremsstrahlung radiation for the energy 
threshold of $10^{12}$ eV so the photons of lower energies are absent in the distribution.
In this way we take into account the majority of the photons even if those of
energies lower than $10^{16}$ eV may have little impact on the longitudinal profile of the
resultant air shower. The total energy fraction carried by such photons is very low 
(see the top right plot in Fig. \ref{20strong_profiles}) and the induced subshowers
fade very quickly. 
In the simulation run presented in Fig. \ref{20strong_profiles},
only 78 out of 1000 primary photons did not convert into $e^+e^-$, these are seen 
in the top right plot as the narrow peak at $E = 10^{20}$~eV.

The distributions of the electrons in the same sample of 1000 preshowers
are shown in the bottom row of Fig. \ref{20strong_profiles}.
Again, the spectrum of particles is plotted to the left and the spectrum
weighted by energy to the right. Only in three cases out of 1000 a bremsstrahlung photon
converted into $e^+e^-$ so the total number of electrons is 1850 and not 1844. When the conversion
happens at a lower altitude, the electrons radiate less bremsstrahlung before 
entering the atmosphere. These cases
are seen in the spectrum as a tail towards high energies.

\begin{figure}
\begin{center}
\includegraphics[width=1.0\textwidth,angle=0]{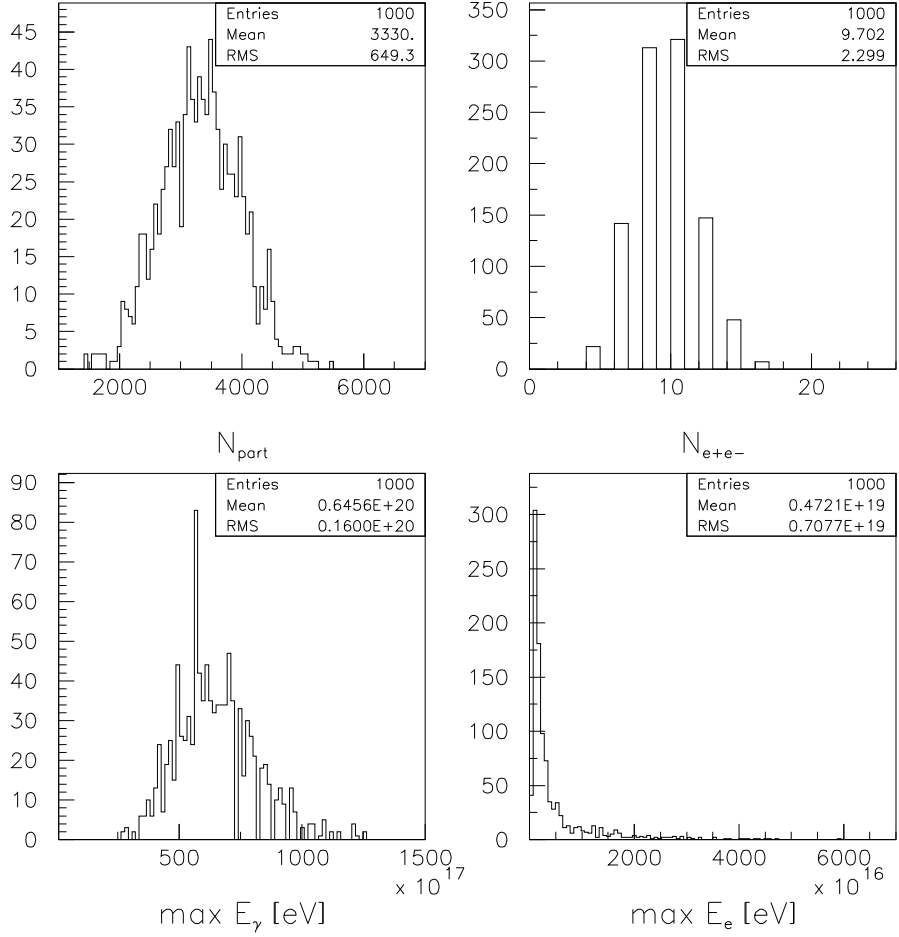}
\end{center}
\caption {
Distributions of $N_{part}$ (top left), $N_{e^+e^-}$ (top right), 
max $E_\gamma$ (bottom left) and max $E_e$ (bottom right) 
for $E_0=10^{21}$ eV and the strong $B_\bot$ arrival direction.}
\label{21strong_detail}
\end{figure}

Some characteristics of the preshowers shown in Figure~\ref{20strong_profiles}
are summarized in Table~\ref{pre_examples} together with some other
exemplary preshower cases.
The conversion probabilities are consistent with those in Figure \ref{maps}.
For higher initial energies gamma conversion takes place at higher altitudes 
(see Fig. \ref{fpp}) and the resultant electrons radiate more bremsstrahlung
photons so the number of particles is larger. 
Additionally for energies as high as $10^{21}$ eV there is more than one
gamma conversion per each preshower which also enlarges $N_{part}$. 
Comparing strong $B_\bot$ and weak $B_\bot$ directions
for a given energy, the large directional dependence of preshower evolution can be noted.
In case of $E_0=10^{20}$ eV from the
weak $B_\bot$ direction we expect unconverted photons initiating cascades with 
a large depth of shower maximum
$X_{max}$ due to the LPM effect. The strong $B_\bot$ forces the initial
energy to be split into about 500 particles of which the highest energy particle
typically
is a photon with the energy about 5 times lower than the initial one. This means
that the longitudinal profile of the resultant air shower will have the maximum
higher in the atmosphere, 
which will make it more similar to a proton-like profile. The energy splitting in this case
will compete with the LPM effect.   
This splitting effect for $E_0=10^{21}$ eV for the strong $B_\bot$ direction is even
more dramatic: the highest energy particle in the preshower is more than 16 times less
energetic than the primary photon.
The maximum $\gamma$ energy in the preshower for the weak field direction is higher
than for the strong field since the conversion probability is lower in a weaker field
(see Fig. \ref{conv_vs_E}). Detailed distributions of <$N_{part}$>, <$N_{e^+e^-}$>, 
<max $E_\gamma$> and <max $E_e$> for $E_0=10^{21}$ eV and the strong $B_\bot$
are shown in the top left, top right, bottom left and
bottom right plots of Figure \ref{21strong_detail}, respectively.

\begin{figure}
\begin{center}
\includegraphics[width=0.7\textwidth,angle=0]{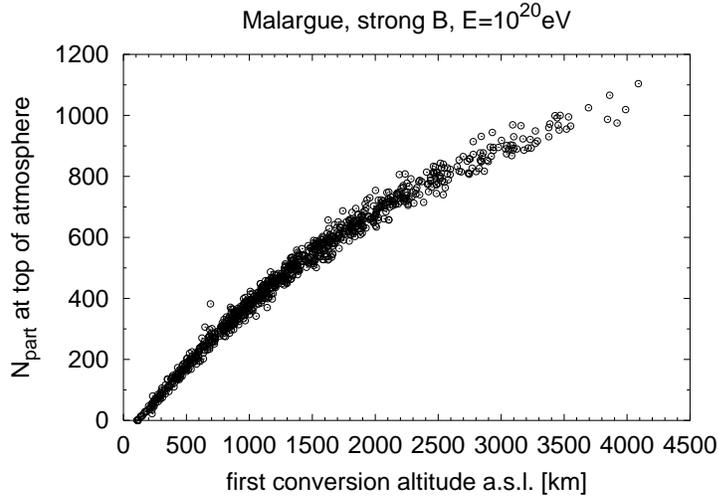}
\end{center}
\caption {Number of particles in the preshower for different altitudes of
first $\gamma$ conversion. Plotted are the preshowers initiated by $10^{20}$ eV photons
arriving from the strong $B_\bot$ direction.}
\label{cor20nfpp}
\end{figure}

As mentioned above, the number of particles in the preshower is connected
to the altitude of the first gamma conversion. This correlation is presented in
Figure \ref{cor20nfpp}.  
The higher altitude of conversion gives more particles at the top of the atmosphere because
the electron-positron pair radiates bremsstrahlung over a longer path. 
The increase in particle number is slightly larger at low altitudes because of the stronger
magnetic field.

\begin{figure}
\begin{center}
\includegraphics[width=0.7\textwidth,angle=0]{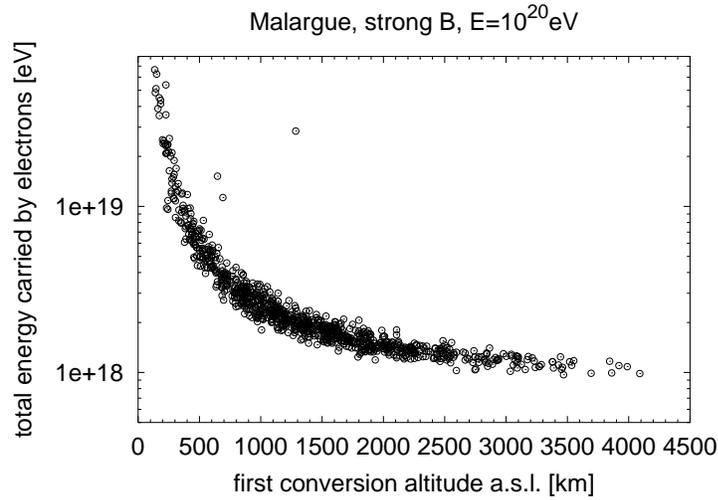}
\end{center}
\caption {Energy carried by the preshower electrons at the top of atmosphere vs
the altitude of the first $\gamma$ conversion for a primary photon energy of
$10^{20}$~eV 
in the strong field direction.
The three points in excess of the general trend are the cases where the primary
photon converted at high altitude and one of the bremsstrahlung photons converted
close to the top of the atmosphere. For further comments see the text.
}
\label{cor20efpp}
\end{figure}

For the same reasons the total energy fraction carried by the electrons should be smaller
for higher conversion altitudes. This dependence is shown in Figure~\ref{cor20efpp}. 
One notes that for high altitudes of the first conversion only about 1\% of the initial energy
of $10^{20}$~eV remains in the electrons at the
top of the atmosphere, while all the rest is radiated
into photons. With decreasing altitudes the energy radiated into photons decreases.
The three points in excess of the general trend in Figure \ref{cor20efpp} 
are the cases where the primary
photon converted at high altitude and one of the resulting bremsstrahlung 
photons converted close to the top of the atmosphere - at 228, 442 and 133 km a.s.l
respectively for the points from right to left. The threshold for $\gamma$
conversion in this case is about $5\cdot10^{19}$ eV, so at least one member of $e^+e^-$
pair had an initial energy greater than $10^{19}$ eV, and since its 
birth altitude was relatively low, there was little time for radiation.   

\subsection{Atmospheric cascading: PRESHOWER and CORSIKA}

The results presented above give a general idea about the characteristics of 
preshowers initiated by UHE photons before entering the atmosphere. The next step is
to simulate air showers induced by different preshowers or unconverted photons
and compare the results with the proton-induced showers in order to investigate
the signatures of UHE photons that will be measurable by the
Pierre Auger Observatory. For this purpose we connected the PRESHOWER program
to CORSIKA, a commonly used air shower simulation code.

The connection between the two codes is 
organized as follows. First, the distribution
of the preshower particles at the top of the atmosphere is simulated with the original
preshower code, then the output is passed to CORSIKA and finally the resulting air shower
is simulated as a superposition of the subshowers initiated by the preshower particles.

Electromagnetic interactions are simulated in CORSIKA using an
adapted version~\cite{upgrade} of the EGS4 code~\cite{egs},
which includes the Landau-Pomeranchuk-Migdal effect~\cite{lpm}.
To reduce the computational effort in CPU time,
   the technique of particle thinning~\cite{hillas},
   including weight limitation \cite{kobal,risseicrc}, is applied.
   More specifically, a thinning level of $10^{-5}$ has been chosen,
   and particle weights are limited to $10^{-5}E_0$/GeV,
   with $E_0$ being the primary energy.
   This keeps artificial fluctuations introduced by the contribution
   of individual particles sufficiently small
   for the analysis of shower maximum depths~\cite{gap-edep}.

In Table \ref{markus_tab} we compare $X_{max}$ and RMS of $X_{max}$ for gamma 
induced showers of different primary energies and arrival directions. 
The $X_{max}$ of proton-induced showers using the high-energy hadronic
interaction model QGSJET~01~\cite{qgsjet} is 
820 (870) g/cm$^2$ for $10^{20}$ ($10^{21}$)~eV, and using the
SIBYLL~2.1~ model \cite{sibyll21} 855 (915) g/cm$^2$.
The RMS in all cases is about 50$-$60 g/cm$^2$.
Both for $E_0=10^{19.5}$~eV in strong $B_\bot$ direction
and for $E_0=10^{20}$ eV in 
weak $B_\bot$ direction, almost every photon remains
unconverted when entering the atmosphere,
resulting in deep $\langle X_{max} \rangle$ and large fluctuations due
to the LPM effect.
A comparison to the longitudinal profile of hadronic primaries shows that 
unconverted primary photons might be well 
distinguishable from primary protons or nuclei even on event-by-event basis.

Almost all the photons at $10^{20}$ eV arriving from the strong field direction convert into
$e^+e^-$ pairs. As the most energetic bremsstrahlung photons have
energies several times lower than the primary photon energy, 
the subshowers develop faster. On average, the showers reach their maxima earlier 
than the one induced by an unconverted photon, but still not as early as hadronic showers.
For larger statistics of 10$^{20}$ eV events and if a substantial fraction
of UHE cosmic rays are photons, one expects to see, as described above, the directional 
anisotropy in $\langle X_{max} \rangle$ and RMS($X_{max}$) and 
on this basis one may draw conclusions on the primary composition.
The absence
of this dependence would allow us to set upper limits on the photon flux.

At the primary energy of $10^{21}$ eV all photons convert, whatever the arrival direction.
We still see the directional dependence of $\langle X_{max} \rangle$ but it is not
as pronounced as previously. The fluctuations of $X_{max}$ in this case 
are significantly lower (by about a factor 4) than for $10^{20}$ eV primaries. 
Table \ref{markus_tab} shows that at $10^{21}$ eV it 
is more difficult to distinguish a photon primary
on the basis of the $X_{max}$ value from a proton one on an event-by-event basis.
It is interesting, however, to note the directional dependence of the elongation rate.
From Table \ref{markus_tab} we see that at Malarg\"ue, for the strong $B_\bot$
direction, the elongation rate of photon-induced showers 
between $10^{20}$ eV and $10^{21}$ eV is much less than
for proton or iron showers. For the weak $B_\bot$ we even have a {\it negative} elongation rate.
This is because the preshowering effect for photons at $10^{21}$ eV splits the initial energy
into energies less than $10^{20}$ eV and at this energy level, for the weak $B_\bot$
direction, almost all the primary photons remain unconverted and they induce air showers
with deeper $X_{max}$. 
Low or negative elongation rates could be an
additional good signature of 
photon showers.

More detailed simulations of the signatures mentioned above will allow us
to quantify the expected sensitivity of the
Auger experiment to a photon fraction in UHE cosmic ray flux. 
Also other observables will be studied, 
especially the muon content and the features of the lateral distribution 
measured by the surface detector, 
since for the surface measurements the event sample will be about 10 times 
more numerous than for the fluorescence detector. 

\section{Summary}
\label{summary}
We presented a method for studying the signatures of UHECR gamma showers
and the practical implementation of this method in the PRESHOWER
program. This program was tested extensively both in stand-alone mode
and as a part of CORSIKA and was checked for consistency with results quoted in other
publications. The preshower option will be accessible in the future
release of CORSIKA. 

With the code working in stand-alone mode, we studied the preshower characteristics
for different primary energies and arrival directions.
The strong directional and energy dependence of the preshower effect requires
a concise treatment in the simulations.
Anisotropies in the preshower cascade translate into anisotropies of air shower
observables such as $\langle X_{max} \rangle$ and RMS($X_{max}$). First results
obtained with PRESHOWER combined with the CORSIKA air shower simulation code
for the conditions of the Auger Observatory confirm the expectation that 
photon showers can be distinguished from hadron ones.
Further analyses including other observables such as the muon shower content 
are in progress.

The new experiments with large collecting apertures
give promising prospects for acquisition of a substantial amount of extensive air
shower data in the UHE range of cosmic-ray spectrum.
An important step towards explaining the origin of UHE cosmic rays will be the
observation of primary UHE photons or specifying stringent upper limits for the photon flux.
This requires a detailed understanding of the UHE photon
shower characteristics.
The presented PRESHOWER code can serve as tool for such simulations.
 
\section{Acknowledgements}

We thank R. Engel, M. Kachelriess, S.S. Ostapchenko for valuable 
remarks and suggestions.

This work was partially supported by the Polish State Committee for
Scientific Research under grant No. PBZ KBN 054/P03/2001 and 2P03B 11024
and by the International Bureau of the BMBF (Germany) under grant
No. POL 99/013.






\newpage
\section*{A Compilation of procedures/functions}
\begin{description}

\item \texttt{preshw(*id, *E\_gamma, *the\_loc, *phi\_loc, *toa, 
*glong, *glat, \\ *igrf\_year, *iiprint, *corsika, *nruns, part\_out[10000][7], \\
*r\_zero, *r\_fpp, *N\_part\_out)}:\\
This is the main procedure of PRESHOWER which simulates the propagation of an 
UHE photon before its entering the    
Earth's atmosphere. It takes into account the effect of $\gamma$         
conversion into $e^{+/-}$ pair and subsequent bremsstrahlung 
radiation of the electrons in the geomagnetic field. As a result a bunch  
(a preshower) of particles is obtained, mainly photons and a few electrons,    
instead of the primary photon. The information about all the        
particles in the preshower is saved (in  
the stand-alone mode) or returned to CORSIKA (PRESHOWER-CORSIKA mode). 
The output parameters \texttt{r\_zero}, 
\texttt{r\_fpp} and \texttt{N\_part\_out} are required
by CORSIKA only -- they are not used by the stand-alone version.\\                            
INPUT:                                                             
\begin{enumerate}
\item \texttt{*id}: primary particle type: always \texttt{id}=1 (a photon)                        
\item  \texttt{E\_gamma}: primary energy [GeV]                
\item  \texttt{*the\_loc, *phi\_loc}: zenith and azimuth angles (in radians) of the      
             shower trajectory given in the local coordinate system                                 
\item  \texttt{*toa}: the top of the atmosphere [cm]                          
\item  \texttt{*glong}: longitude position of experiment [deg]          
             (Greenwich \texttt{glong} = $0^\circ$, eastward is positive)                 
\item  \texttt{*glat}:  latitude position of experiment [deg]          
             (North Pole: \texttt{glat}=$90^\circ$, South Pole: \texttt{glat}=$-90^\circ$)                   
\item  \texttt{*igrf\_year}: the year in which the magnetic field is to be  
             computed -- this parameter is an input for the \texttt{igrf}      
             procedure                                           
\item  \texttt{*iiprint}: print flag (adjustable only in PRESHOWER-CORSIKA, in
the stand-alone version \texttt{iiprint=1}) \\= 0: suppress printing from \texttt{preshw}; \\
                           = 1: enable printing from \texttt{preshw}       
\item  \texttt{*corsika}: mode flag \\= 1: PRESHOWER-CORSIKA: -- random        
                     generator from CORSIKA is used (\texttt{rndm});\\         
                  = 0: stand-alone version: random generator from Numerical Recipes (\texttt{ran2}) is used                          
\item  \texttt{*nruns}: number of runs                                    
\end{enumerate}
OUTPUT:
\begin{enumerate} 
\item  \texttt{part\_out[10000][7]}: the array to store the preshower particle data, maximum number
of particles: 10000 \\                                              
  part\_out[n][0]: particle type: 1 -- photon, 2 -- $e^+$, 3 -- $e^-$\\                  
  part\_out[n][1]: particle energy [GeV] \\                
  part\_out[n][2-6]: not used                     
\item  \texttt{*r\_zero}: height above sea level of the simulation     
             starting point given in [km] above sea level            
\item  \texttt{*r\_fpp}: height of the first pair production    
             (0 for surviving photons) given in [km] above sea level                
\item  \texttt{*N\_part\_out}: number of preshower particles at the top of the  
             atmosphere (number of entries in \texttt{part\_out} array)       
\end{enumerate}

\item \texttt{bessik(x, xnu, *ri, *rk, *rip, *rkp)}:\\
for argument \texttt{x} computes modified Bessel function \texttt{rk} of fractional order 
\texttt{xnu}; \cite{numrec}

\item \texttt{getrand(seed, mode)}:\\
for a given seed \texttt{seed} returns a random number from interval (0,1) 
choosing an appropriate generator according to the \texttt{mode} flag: 
\texttt{ran2} for the stand-alone mode (\texttt{mode}=0) and \texttt{rmmard} for the
CORSIKA-PRESHOWER (\texttt{mode}=1)

\item \texttt{float kappa(x)}:\\
an auxiliary function used in the bremsstrahlung procedure \texttt{brem}

\item \texttt{brem(B\_tr, E\_e, E\_sg)}:\\
computes bremsstrahlung radiation for the transverse magnetic field \texttt{B\_tr},
electron energy \texttt{E\_e} and bremsstrahlung photon energy \texttt{E\_sg}

\item \texttt{B\_transverse(B, tn)}:\\
getting the geomagnetic field \texttt{B} component which is perpendicular 
to the preshower trajectory \texttt{tn}

\item \texttt{ran2(seed)}:\\
the random number generator used only in the stand-alone mode; \cite{numrec}

\item \texttt{cross(a, b, c)}:\\
returns cross product \texttt{c} of vectors \texttt{a} and \texttt{b}

\item \texttt{dot(a, b)}:\\
returns dot product of vectors \texttt{a} and \texttt{b}

\item \texttt{norm(a, n)}:\\
the input vector \texttt{a} is normalized and returned as vector \texttt{n} 

\item \texttt{normv(a)}:\\
returns the length of vector \texttt{a}

\item \texttt{glob2loc(glob, loc, theta, phi)}:\\
converts global cartesian coordinates \texttt{glob} to the local ones (\texttt{loc})
for the geographical position \texttt{theta} (colatitude) and \texttt{phi} (longitude)

\item \texttt{sph2car(sph, car)}:\\
converts spherical coordinates \texttt{sph} to the cartesian ones \texttt{car}.

\item \texttt{car2sph(car, sph)}:\\
inverse to \texttt{sph2car}

\item \texttt{locsphC2glob(locsph, glob, theta, phi, sing, cosgm, sitec)}:\\
converts local spherical coordinates \texttt{locsph} (CORSIKA frame of reference)
to the global cartesian coordinates \texttt{glob}, using geographical coordinates of the
observatory (\texttt{theta} - colatitude, \texttt{phi} - longitude), global cartesian
coordinates of the observatory (\texttt{sitec}) and sine and cosine of the declination angle
(\texttt{sing}, \texttt{cosg})

\item \texttt{ppp(efrac, B\_tr, E\_g0)}:\\
this auxiliary function is used within \texttt{ppfrac}

\item \texttt{ppfrac(s, mode, B\_tr, E\_g0)}:\\
determines the energy partition after gamma conversion, the computations are performed according
to Ref. \cite{ppdaugherty}; \texttt{s} - seed, \texttt{mode} - determines the type of the 
random number 
generator, \texttt{B\_tr} - transverse magnetic field, \texttt{E\_g0} - primary photon energy

\item \texttt{igrf\_(year, n, r, theta, phi, Br, Btheta, Bphi)}:\\
external Fortran procedure written by Tsyganenko \cite{tsygan}; computes magnetic field 
according to the IGRF model; used within \texttt{b\_igrf}; \texttt{year} - 
the year of the experiment, \texttt{n} - the highest order of the spherical harmonics in
the scalar potential, \texttt{r, theta, phi} - spherical coordinates of the point at which
the field is to be found,
\texttt{Br, Btheta, Bphi} - spherical coordinates of the magnetic field at given position 

\item \texttt{b\_igrf(year, r, theta, phi, bcar)}:\\
for a given \texttt{year} and position (\texttt{r, theta, phi}) computes the 
magnetic field \texttt{bcar} in global cartesian coordinates according to the IGRF model

\item \texttt{bsph2car(colat, longit, bsph, bcar)}:\\
converts the magnetic field from spherical coordinates (\texttt{bsph}) into 
cartesian ones \texttt{bcar}; used only in \texttt{b\_igrf}

\item \texttt{rndm()}:\\
calls CORSIKA random number generator, used only in the PRESHOWER-CORSIKA mode  

\end{description}

\newpage 

\begin{table}
\begin{center}
\caption {The magnetic pair production function $T(\chi)$. The values below differ from the
ones given in Ref.~\cite{Erber}. 
The computation of the limiting cases of $T(\chi)$ (Eq.~(\ref{tlimits})) or checking the
{\it K} function values
in mathematical tables proves that the numbers given here are correct.}
\begin{tabular}{c|ccccccc}
\hline \hline
$\chi$ & 0.2 & 0.4 & 1.0 & 5.0 & 10.0 & 30.0 & 100.0 \\
\hline
$T(\chi)$ & 5.9e-4 & 1.6e-2 & 9.9e-2 & 2.2e-1 & 2.1e-1 & 1.9e-1 & 1.3e-1 \\
\hline
\end{tabular}
\label{Ttab}
\vspace{2cm}

\end{center}
\end{table}

\newpage

\begin{table}
\begin{center}
\caption {
Exemplary preshowers for Malarg\"ue magnetic conditions.
For each type of the preshower determined by the initial
photon energy $E_0$ and the arrival
direction, in columns 3-7 respectively shown are: the fraction of
converting primary photons and
the average values ($\pm$ RMS) for the number of particles, the number of electrons, the
maximum photon energy and the maximum electron energy. Only converted cases were
taken for averaging.
}
\begin{tabular}
{p{0.8cm}p{1.7cm}p{2.0cm}p{1.6cm}p{1.7cm}p{1.6cm}p{1.6cm}} \hline \hline
$E_0$ [EeV]& direction & fraction of converted & $\langle N_{part}\rangle$ & $\langle N_{e^+e^-} \rangle$
& $\langle max E_\gamma \rangle$ [EeV]& $\langle max E_e \rangle$ [EeV]\\ \hline 
50 & strong $B_\bot$
 & 137/1000 &263$\pm$189 & 2.0$\pm$0.0 & 6.8$\pm$3.2 & 4.9$\pm$6.5 \\ 
100 & strong $B_\bot$
 & 922/1000 & 482$\pm$231 & 2.01$\pm$0.11 & 18$\pm$8 & 2.2$\pm$5.3 \\ 
 & weak $B_\bot$
 & 0/1000 & 1 & 0 & 100 & 0 \\ 
1000 & strong $B_\bot$
 & 1000/1000 & 3330$\pm$649 & 9.7$\pm$2.3 & 65$\pm$16 & 4.7$\pm$7.1 \\ 
 & weak $B_\bot$
 & 1000/1000 & 544$\pm$143 & 2.7$\pm$1.0 & 167$\pm$66 & 7.2$\pm$8.3 \\ \hline \hline
\end{tabular}
\label{pre_examples}
\end{center}
\end{table}

\begin{table}
\begin{center}
\caption {$X_{max}$ and RMS values for gamma 
induced showers of two different primary energies and arrival directions. Here the 
strong $B_\bot$ direction
is defined as $\theta=53^{\circ}$, $\phi=177^{\circ}$ and 
weak $B_\bot$ as $\theta=53^{\circ}$, $\phi=357^{\circ}$.}
\begin{tabular}
{p{1.5cm}p{2.5cm}p{2.5cm}p{3cm}p{3cm}} \hline \hline
$E_0$ [eV]& direction & fraction of converted & $\langle X_{max} \rangle$ [g/cm$^2$] & 
$\langle RMS(X_{max}) \rangle$ [g/cm$^2$] \\ \hline 
10$^{19.5}$ & strong $B_\bot$ & 1/50 & 1065 & 90 \\ 
10$^{20.0}$ & weak $B_\bot$ & 1/100 & 1225 & 175 \\ 
                 & strong $B_\bot$ & 91/100 & 940 & 85 \\ 
10$^{21.0}$ & weak $B_\bot$ & 100/100 & 1040 & 40 \\ 
                 & strong $B_\bot$ & 100/100 & 965 & 20 \\ \hline \hline
\end{tabular}\\
\end{center}
\label{markus_tab}
\vspace{5cm}
\end{table}

\newpage



\end{document}